\documentclass[doublecol]{epl2}

\usepackage{amsmath}
\usepackage{amssymb}
\usepackage{graphicx}
\usepackage{color}

\title{Detection of X-ray galaxy clusters based on the Kolmogorov method\footnote{Based on observations obtained with XMM-Newton, an ESA science mission with instruments and contributions directly funded by ESA Member States and NASA}}
\shorttitle{Detection of X-ray galaxy clusters}

\author{V.G.~Gurzadyan$^1$ \and F.Durret$^{2,3}$ \and T.~Ghahramanyan$^1$ \and 
A.L.~Kashin$^1$ \and H.G.~Khachatryan$^1$ \and E.~Poghosian$^1$ 
}

\shortauthor{V.G.Gurzadyan \etal}

\institute{1. Alikhanian National Laboratory and Yerevan State University, Yerevan, Armenia; 
2. UPMC Universit\'e Paris 06, UMR~7095; 3. Institut d'Astrophysique de Paris, 
F-75014, Paris, France
}


\abstract{The detection of clusters of galaxies in large surveys
  plays an important part in extragalactic astronomy, and particularly
  in cosmology, since cluster counts can give strong constraints on
  cosmological parameters.  X-ray imaging is in particular a reliable
  means to discover new clusters, and large X-ray surveys are now
  available. Considering XMM-Newton data for a sample of 40 Abell
  clusters, we show that their analysis with a Kolmogorov distribution
  can provide a distinctive signature for galaxy clusters. The
  Kolmogorov method is sensitive to the correlations in the cluster
  X-ray properties and can therefore be used for their identification,
  thus allowing to search reliably for clusters in a simple way.
}

\begin{document}

\maketitle



\section{Introduction}

The reliable determination of galaxy clusters as entities is crucial
not only for the studies of their individual structure and evolution,
but in the broader context of the formation of large scale structures
in the universe. Various methods to define the cluster membership of
galaxies have been developed, usually based on limited information
such as line-of-sight velocities or photometric redshifts (1D), and
coordinates (2D).  X-ray imaging has provided an important means to detect
clusters as diffuse X-ray sources and to define cluster properties by
allowing to derive their overall gravitational potential, including
dark matter, under the assumption that the X-ray emitting gas is in
hydrostatic equilibrium in the cluster gravitational potential well
dominated by dark matter.

In the present paper we apply the Kolmogorov stochasticity parameter
method and distribution \cite{Kolm,Arnold} to analyze X-ray data, in
order to reveal possible signatures which can allow to identify a
galaxy cluster. This approach quantifies the degree of randomness for
given  sequences of numbers and has been applied to various systems
appearing in number theory \cite{Arnold_UMN,Arnold_MMS}. We have also applied
this method to the cosmic microwave background (CMB) radiation signal
\cite{GK2008,G2011}. It allowed to identify non-Gaussianities in the
CMB such as the Cold Spot in the maps of the Wilkinson Microwave
Anisotropy Probe (WMAP). The behavior of the stochasticity parameter
supports the void nature of that anomaly, and another non-Gaussianity,
the North Cold spot, has also been identified in the north sky by the
same method \cite{GK2009}. This approach was also efficient
at detecting point sources in the WMAP maps, including the
prediction \cite{G2010} of gamma-ray sources (quasars, blazars)
which have later been identified by the Fermi satellite \cite{Fermi1}.

Much has been learnt these past ten years on the properties of X-ray
clusters, based on the data acquired by the XMM-Newton, Chandra and
Suzaku satellites, and the corresponding references are too numerous
to be quoted here (for overview see e.g. NASA and ESA sites for X-ray
missions {\bf http://www.nasa.gov/missions/index.html; http://www.esa.int/esaSC}). 
In particular, with its large collecting area,
XMM-Newton has allowed not only to draw accurate emissivity maps, but
also to compute hardness ratio, temperature and metallicity maps of
the X-ray gas (as first obtained by
\cite{Finoguenov04,Durret05}). These maps give a much deeper insight
on the cluster properties, in particular on the cluster merging
history. They have shown unambiguously that even clusters with X-ray
emissivity maps showing a relatively relaxed appearance could in fact
be undergoing one or several mergers (see e.g. \cite{Durret08} and
references therein). Several cluster surveys based on XMM-Newton data
are available, the first being those of
\cite{Basilakos04,Schwope04,Valtchanov04}. The XMM-Newton archive is
now extremely large, and a simple method applied to detect clusters
throughout this archive would be a very powerful tool to obtain a
large and homogeneous sample of clusters. Note that presently very
large and homogeneously detected cluster catalogs are also being
obtained, based on the Sunyaev-Zel'dovich effect, both with ground
based facilities such as the South Pole Telescope
\cite{Williamson11} or the Planck satellite \cite{Aghanim11}.

The aim of the present study is not to analyse the properties of
individual clusters, but to show that the Kolmogorov descriptor,
considered as a quantifier of correlations in datasets, can be used
for the identification of galaxy clusters based on X-ray data. For
this particular aim we will use a sample of Abell clusters with
available XMM-Newton data.

The paper is structured as follows: we briefly describe our method in
Section~2, we present our sample of X-ray clusters and analysis
together with our results in Section~3. Conclusions are drawn in
Section~4.

\section{The method}

The definition of the Kolmogorov stochasticity parameter
\cite{Kolm,Arnold} is given for a finite random sequence of real
numbers $x_1, x_2, \ldots , x_n$ so that the values of $x_n$ are
sorted in an increasing manner: $x_1 \leq x_2 \leq \ldots \leq x_n$.

One can define an empirical distribution function:
$$C_n(X)= (number\, of\, the\, elements\, x_i \leq  X),$$
where

\begin{equation}
C_n(X)= \left\{
\begin{array}{rl}
	0, & X < x_1 \\
	k / n, & x_k \leq X < x_{k+1} \\
	1, & x_n \leq X\\
\end{array}
\right.
\label{eq:empiricdistribution}
\end{equation}

The theoretical distribution  function $C(X)$ is then:
$$C(X) = n \cdot(probability\, of\, the\, event\, x \leq X).$$

The stochasticity parameter $\lambda_n$  for a sequence  of $n$ values of $x$ is defined as:
\begin{equation}
\lambda_n =  sup_X|C(X)-C_n(X)|/ \sqrt{n}.\\
\label{eq:lambda}
\end{equation}

According to Kolmogorov's theorem, $\lambda_n$ is a random number with
an empirical distribution function
$$\Phi_n(\Lambda) = (number\, of\,  elements\, \lambda_n \leq  \Lambda),$$
uniformly converging to $\Phi(\Lambda)$ at $n \rightarrow \infty$ for
any continuous distribution $C(X)$
\begin{equation}
\Phi_n(\Lambda) \rightarrow \Phi(\Lambda),
\label{eq:philimit}
\end{equation}
where 
\begin{equation}
\Phi(\Lambda) = \sum_{k=-\infty}^{+\infty}{(-1)^k e^{-2k^2\Lambda^2}},\\
\ -\infty <  k < +\infty.
\label{eq:phi}
\end{equation}

The distribution function $\Phi(\Lambda)$ varies monotonically from
$\Phi(0)=0$ to $\Phi(\infty)=1$.

According to the definition, for large enough $n$ and random sequence
$x_n$ the Kolmogorov stochasticity parameter $\lambda_n$ will have a
distribution close to $\Phi(\Lambda)$. If the sequence is not random,
the distribution will be different. So $\Phi(\Lambda)$ denotes a
quantitative measure for the randomness of a sequence. Arnold  has
shown the informativity of this method already for sequences of n=15 \cite{Arnold}.

\section{Analysis and results}

Our analysis is based on a sample of 40 Abell clusters with data
available in the second XMM-Newton Serendipitous Source Catalogue,
2XMM, released on April 15, 2010: the 2XMMi-DR3
catalogue\footnote{http://xmmssc-www.star.le.ac.uk/Catalogue/2XMM/} in
the [0.2-12]~keV band, of median flux $2.5\, 10^{-14}$ erg\,
cm$^{-2}$\, s$^{-1}$.  The X-ray images of the clusters
represent datasets of X-ray fluxes for pixels with given
coordinates. Table~1 shows the relevant data for the 40 clusters;
the first 17 have redshifts lower than 0.03, although no noticable
redshift dependence have been noticed at the study below.  
For illustration in {\bf Figure~1 we represent the X-ray image} for a particular cluster, A2870, and in Figure 2
the cuts of its X-ray flux distribution over the Galactic {\it l}
and {\it b} coordinates.

We then create Gaussian mocks of the clusters, {\bf namely, the mock of a given cluster
is obtained as a Gaussian distribution of the flux with the median and standard deviation defined
by the real data of its X-ray pixelized distribution};
the cuts along two axes are also shown in Figure~2. Since
the Gaussian mocks of the clusters cannot possess the information on
the cluster potential carried by the real X-ray images, they can serve as
a calibration for the correlations in the real data, i.e. what we have
to quantify using Kolmogorov's method. Comparing the plots in the left
and right columns in Figure~1, we see that although the real cluster
data and the Gaussian ones exhibit certain discrepancies, they do
not differ radically, and in any case they cannot be distinguished
reliably only via such a comparison.

\begin{table}
  \caption{Sample definition for the 40 clusters: running number, cluster 
    name, redshift, galactic coordinates (of centers),  
    number of  pixels for which X-ray fluxes are given; 
    the first 17 clusters have redshifts 
    lower than 0.03, the others have a higher redshift.}
\renewcommand{\baselinestretch}{1.00}
\renewcommand{\tabcolsep}{2.5mm}
\small
\begin{center}
\begin{tabular}{rrcrrr}
\hline
\hline
 N & Abell &   z    & l, deg & b, deg & Pixel\\
\hline
 1 & A0194 & 0.0178 & 141.95 & -63.00 &   97 \\
 2 & A0262 & 0.0151 & 136.59 & -63.00 &  183 \\
 3 & A0400 & 0.0232 & 170.53 & -44.90 &  116 \\
 4 & A1060 & 0.0114 & 269.64 &  26.51 &   85 \\
 5 & A1314 & 0.0323 & 151.84 &  63.57 &   86 \\
 6 & A1367 & 0.0208 & 234.81 &  73.03 &  102 \\
 7 & A1656 & 0.0219 &  58.09 &  87.96 &  758 \\
 8 & A2052 & 0.0338 &   9.40 &  50.10 &  324 \\
 9 & A2063 & 0.0341 &  12.86 &  49.81 &   99 \\
10 & A2147 & 0.0338 &  28.81 &  44.49 &   45 \\
11 & A2197 & 0.0296 &  64.82 &  43.80 &   45 \\
12 & A2199 & 0.0309 &  62.90 &  43.70 &  103 \\
13 & A2870 & 0.0225 & 294.81 & -69.95 &   73 \\
14 & A2877 & 0.0235 & 293.14 & -70.88 &  169 \\
15 & A3526 & 0.0102 & 302.42 &  21.56 &  248 \\
16 & A3574 & 0.0148 & 317.47 &  30.94 &  162 \\
17 & A3581 & 0.0218 & 323.13 &  32.85 &  285 \\
\hline
18 & A0085 & 0.0543 & 115.06 & -72.06 &   83 \\
19 & A0087 & 0.0538 & 116.01 & -72.55 &  100 \\
20 & A0119 & 0.0430 & 125.76 & -64.11 &  159 \\
21 & A0133 & 0.0554 & 149.10 & -84.09 &   77 \\
22 & A0150 & 0.0576 & 129.62 & -49.47 &   60 \\
23 & A0152 & 0.0569 & 129.71 & -48.65 &   65 \\
24 & A0168 & 0.0438 & 136.67 & -62.04 &  127 \\
25 & A0222 & 0.2110 & 162.67 & -72.20 &  217 \\
26 & A0223 & 0.2070 & 162.42 & -72.00 &  217 \\
27 & A0576 & 0.0381 & 161.42 &  26.24 &  315 \\
28 & A1213 & 0.0468 & 201.46 &  68.99 &   94 \\
29 & A1775 & 0.0705 &  31.93 &  78.71 &   73 \\
30 & A1795 & 0.0619 &  33.79 &  77.16 &  265 \\
31 & A1983 & 0.0424 &  18.95 &  60.12 &  113 \\
32 & A2065 & 0.0714 &  42.88 &  56.56 &   98 \\
33 & A2151 & 0.0354 &  31.60 &  44.52 &  112 \\
34 & A2163 & 0.2005 &   6.76 &  30.52 &  143 \\
35 & A2717 & 0.0478 & 349.21 & -76.48 &  145 \\
36 & A3112 & 0.0738 & 252.94 & -56.08 &   88 \\
37 & A3128 & 0.0587 & 264.74 & -51.11 &  111 \\
38 & A3158 & 0.0585 & 256.06 & -48.92 &   78 \\
39 & A3266 & 0.0577 & 272.19 & -40.15 &  393 \\
40 & A4059 & 0.0463 & 356.84 & -76.06 &  113 \\
\hline
\end{tabular}
\end{center}
\end{table}

We then apply the Kolmogorov technique to the X-ray data for each
cluster. Namely, we estimate the Kolmogorov stochasticity parameter
$\lambda$ and the function $\Phi(\lambda)$ for the X-ray images, and
do the same for the Gaussian images of each cluster. The maximal and minimal values of
the resulting $\Phi$ functions are shown in Fig.~2 for all 40
clusters, as well as for their Gaussian mock clusters.  We see that for real
X-ray images of clusters the function $\Phi$ outside the center
remains~1. On the other hand, its
behavior for the Gaussian images is entirely different, i.e. far smaller
at all radii. To test the dependence on the number of pixels in Fig.~3 we show 
these distributions for 4 clusters with large scatter in the pixel numbers:
A0150 - 60; A3526 - 248; A0576 - 315; A2052 - 324. It is clear that, the higher
is number of pixels, more outlined is the difference between the real clusters
and the mock Gaussian ones.

\section{Conclusions}

We analyzed a sample of 40 Abell X-ray clusters with available 
XMM-Newton data. By applying the Kolmogorov method to the X-ray data
and to Gaussian simulated images, we revealed a clear difference in
the behavior of Kolmogorov's function $\Phi$ for real clusters and for
their Gaussian simulated images.  The X-ray image of a real galaxy
cluster carries correlations reflecting the gravitational potential of
the cluster and hence, the physical conditions of the hot gas
responsible for the X-ray emission.  Such correlations are not present
in the randomly redistributed simulated Gaussian clusters. Although 
the original X-ray clusters and their simulated
Gaussian versions do not show radically different behaviors in their
flux distributions, a difference, however, is clearly outlined
quantitatively in the behavior of the Kolmogorov function $\Phi$. For
the real X-ray images of the clusters the function $\Phi$ tends to 1 
outside the cluster cores,
while for the Gaussian mock clusters it remains smaller at all radii.

Thus, the Kolmogorov method applied to X-ray images enables to
distinguish a galaxy cluster from a random X-ray flux distribution,
thus opening new possibilities in the detection of galaxy clusters.

\begin{figure}
\begin{center}
\includegraphics[width=0.3\textwidth]{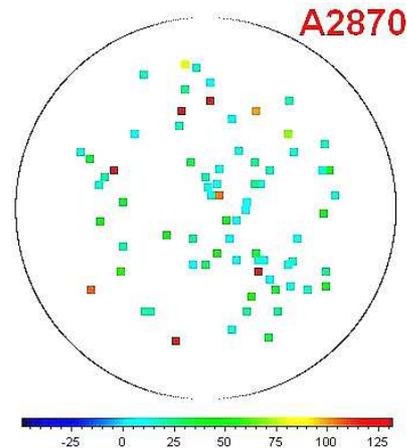} 
\caption{X-ray image of the cluster Abell 2870 obtained by
   XMM-Newton. The circle has a radius of 0.3$^{\circ}$; the '0' of the
   scale bar below the figure corresponds to a flux of 41.3 in units of
   $10^{-15}$ erg\, cm$^{-2}$\, s$^{-1}$, the minimal and maximal
   values of the bar are 7.3 and 405.6, respectively, in the same units.}
\end{center}
\end{figure}

\begin{figure}
\begin{center}
\includegraphics[width=0.48\textwidth]{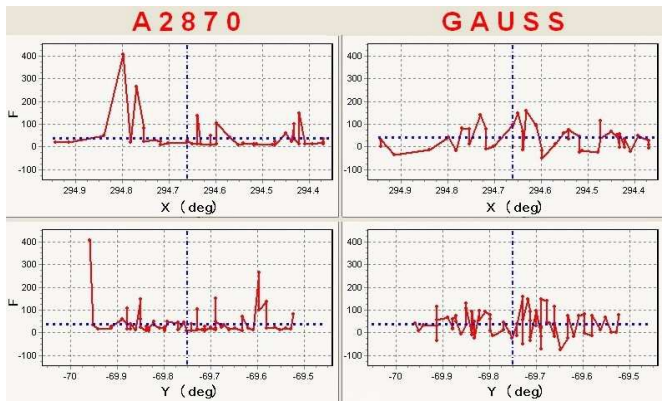} 
\caption{Distribution of the pixel X-ray flux along the {\it l} (top)
  and {\it b} (bottom) axes for A2870 (left) and for the simulated
  Gaussian distribution of the X-ray data parameters (right).  }
\end{center}
\end{figure}

\begin{figure}
\begin{center}
\includegraphics[width=0.45\textwidth]{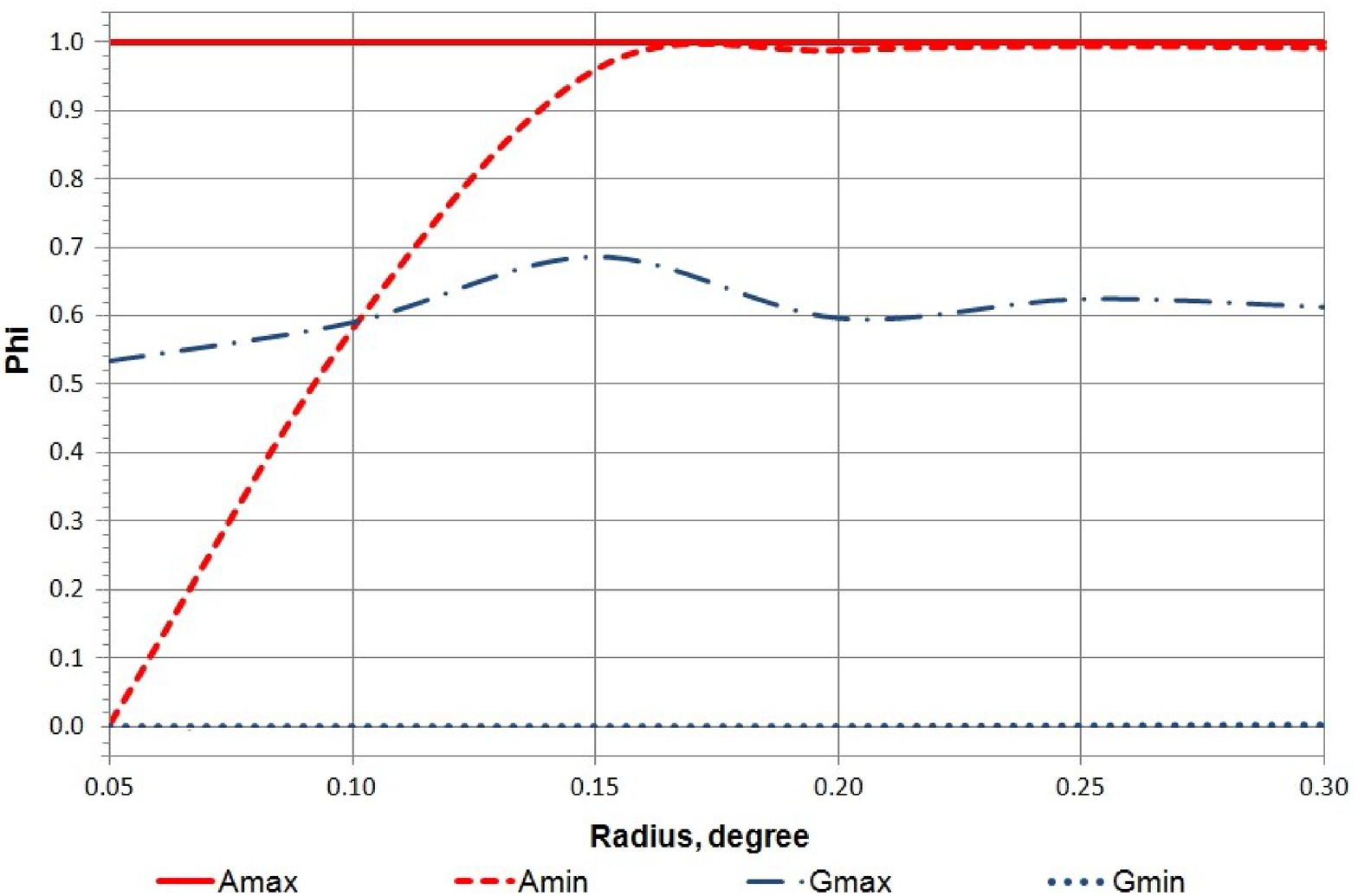} 
\caption{Maximal and minimal values of the Kolmogorov function $\Phi$ vs the cluster
  radius (in degrees) for X-ray data for 40 galaxy clusters, as well as the same for
  corresponding Gaussian mock clusters (in blue).}
\end{center}
\end{figure}

\begin{figure}
\begin{center}
\includegraphics[width=0.45\textwidth]{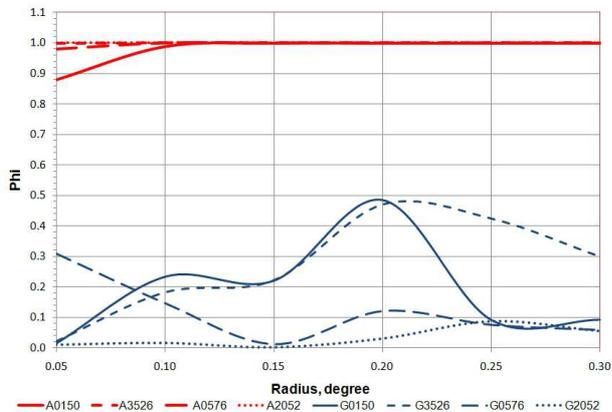} 
\caption{Kolmogorov function vs the radius for 4 clusters with large scatter in the pixel numbers
   (along with their Gaussian versions):
   A0150 - 60; A3526 - 248; A0576 - 315; A2052 - 324.}
\end{center}
\end{figure}

\end{document}